\documentclass[nofootinbib]{revtex4}
\usepackage{graphicx}
\usepackage{url}
\usepackage{latexsym}
\begin{document}
\input{epsf.sty}

\def\affilmrk#1{$^{#1}$}
\def\affilmk#1#2{$^{#1}$#2;}
\def\be{\begin{equation}}
\def\ee{\end{equation}}
\def\bea{\begin{eqnarray}}
\def\eea{\end{eqnarray}}

\title{Perturbations of the Quintom Models of Dark Energy and
the Effects on Observations}

\author{Gong-Bo Zhao$^1$, Jun-Qing Xia$^1$, Mingzhe Li$^2$,
Bo Feng$^1$, and Xinmin Zhang$^1$}
\affiliation{$^1$ Institute of High Energy Physics, Chinese Academy
of Sciences, P.O. Box 918-4, Beijing 100049, People's Republic of
China} \affiliation{$^2$ Institut f\"{u}r Theoretische Physik,
Philosophenweg 16, 69120 Heidelberg, Germany}

\begin{abstract}
We study in this paper the perturbations of the quintom dark
energy model and the effects of quintom perturbations on the
current observations. Quintom describes a scenario of dark energy
where the equation of state gets across the cosmological constant
boundary $w = -1$ during evolution. We present a new method to
show that the conventional dark energy models based on single k-essence field
and perfect fluid cannot act as
quintom due to the singularities and classical
instabilities of perturbations around  $w = -1$. One
needs to add extra degrees of freedom for successful quintom
model buildings. There are no singularities or classical
instabilities in perturbations of realistic quintom models and
they are potentially distinguishable from the cosmological
constant. Basing on the realistic quintom models in this paper we
provide one way to include the perturbations for dark energy
models with parametrized equation of state across $-1$. Compare
with those assuming no dark energy perturbations, we find that the
parameter space which allows the equation of state to get across
$-1$ will be enlarged in general when including the perturbations.
\end{abstract}


\maketitle


\setcounter{footnote}{0}

\section{Introduction}


In 1998 two groups \cite{Riess98,Perl99} have independently
discovered the accelerating expansion of our current universe based
on the analysis of Type Ia Supernovae (SN) observations of the
redshift-distance relations. In the framework of
Friedmann-Robertson-Walker (FRW) cosmology, the acceleration has
been attributed to the mysterious source dubbed dark energy. The
simplest candidate for dark energy is a small positive cosmological
constant, but it suffers from the difficulties associated with the fine
tuning and the coincidence problem \cite{SW89,ZWS99}. The most
popular alternative to the cosmological constant is the model of
rolling scalar field--quintessence \cite{pquint,quint}. In most
cases the quintessence equation of state (EOS) $w$ changes slowly
with time and can be well approximated with a constant $w$ with
$w\ge -1$ \cite{HWDCS99,WCOS2000}. In the early probes of new
physics, cosmologists have assumed a cosmological constant as the new
component \cite{CPT92,KT95,OS95,Riess98,Perl99} and later
fitted directly to the dynamical quintessence models
\cite{CDF96,ZWS99}, or used a constant $w$
\cite{CDS98,GSST98,HWDCS99,WCOS2000} where $w$ was restricted in the region of $w\ge -1$. In
Ref.\cite{phant99} the author firstly extended the fitting of dark
energy to include $w<-1$ and found some mild preferences. The author
constructed a toy model of rolling scalar field with a negative
kinetic term and called it phantom \cite{phant99}. The model of
phantom has some theoretical problems \cite{Phtproblms} and there
have been many attempts towards resolving them \cite{SPhtproblms}.

The accumulation of the observational data
\cite{Bennett03,Tonry03,Knop03,Tegmark03,Riess04,Allen04} has opened
a robust window for probing the more detailed behaviors of dark
energy. There have been many studies in reconstructing the evolution
of its energy density \cite{wangy04} or equation of state
\cite{sahni00,Huterer,dd03} as a function of the redshift. Various
parametrizations of $w$ as well as dark energy models have also
been considered to fit directly to the observational data (e.g.
\cite{DES,DES1,DES2,sahni,cooray,ex,ex1,quintom,seljak04,tao05}).
Based on the fact that current observations cannot exclude dark
energy models with the equation of state getting across $-1$ during
evolution with the redshift, we proposed a model dubbed
quintom\cite{quintom}.
 The model of quintom is a
new scenario in the sense that the conventional quintessence or
phantom models cannot realize the crossing of the cosmological
boundary. Along this line the author in Ref.\cite{Vikman04} has
demonstrated that in the framework of general relativity the model
of k-essence \cite{kessence}, where the scalar field of dark energy
has non-canonical kinetic terms, cannot realize such a crossing
behavior. A toy model with two rolling scalar fields
which have opposite kinetic energy terms can easily realize the
transition and it can be regarded as the simplest quintom model
\cite{quintom,quintom1}. Recently a single field quintom model was proposed in \cite{MFZ05} by
adding higher derivative operators in the
Lagrangian. In the simplest case such a model is equivalent to the two-field case as proposed in \cite{quintom}.
In addition, the quintom model of dark energy are
different from the quintessence or the phantom in the
determination of the evolution and the fate of the universe.
Due to its distinctive properties,
the quintom model with oscillating equation of state across $-1$ can lead to the oscillations of the Hubble constant
and a new scenairo of recurrent universe \cite{FLPZ}, which to some extent unifies the early
inflation \cite{Pinflation} and the current acceleration of the
universe. Recently there have been a lot of interests in the
phenomenological studies relevant to quintom models in the
literature \cite{wei,zhang,hu,xia,michael,li,xfzhang,relevnt,relevnt1,NOT05,0504518}.

Current supernovae data alone, which make the only direct detection
of dark energy, seem to favor a quintom-like model at around
2-$\sigma$ level \cite{cooray,sahni,quintom}. The quintom model is
also mildly favored in the combined analysis with the cosmic
microwave background (CMB), large scale structure (LSS) and
supernovae data \cite{ex,ex1}. However, when some other
observational data sets (such as the new observational data based
on Chandra measurements of the X-ray gas mass fraction in 26 X-ray
luminous galaxy clusters \cite{Allen04} or the recent new
constraints from the bias and Ly$\alpha$ forest of the Sloan
Digital Sky Survey (SDSS)) have been taken into account the
situation changes and the preference for quintom-like dark energy
models becomes weak \cite{quintom,seljak04}. However the previous
fittings in the literature on quintom-like dark energy models have
either fully or partially neglected the perturbations, which in
some sense do not describe the realistic models with EOS across $-1$
and will lead to some bias in the fittings. The aim of this paper
is to develop a self-consistent way to include the perturbations
of quintom in light of the observations. We will present a simple
new method to show that conventional single perfect fluid and k-essence dark energy
models cannot act as quintom, which is due to the singularities and classical
instabilities of perturbations. Based on the realistic quintom models
in this paper we will provide one way to include the perturbations for dark energy models with
parametrized equation of state across $-1$. Compared with those assuming no dark energy
perturbations, we find that when including the
perturbations the parameter space which allows the equation of state to get
across $-1$ will be enlarged in general.

This paper is organized as follows: in section II we discuss the difficulty of quintom
model buildings and provide a new proof regarding the impossibility of single perfect fluid and k-essence model as quintom,
then present some viable quintom models; in section III we study in detail the perturbations of the quintom models;
in section IV we investigate the possible
signatures of quintom models of scalar fields and the effects of
quintom perturbations on the observations; in section V  we provide one way to
include the perturbations for models of dark energy with a
parametrized equation of state across $-1$; we conclude in section VI.


\section{Quintom Model Building}
\subsection{Difficulties of Quintom Model Buildings}

We start with a brief overview on the arguments against the
possibility of realizing the quintom scenario with a single fluid or
a single scalar field in the conventional framework.

Consider firstly a single perfect fluid, the energy-momentum tensor
has the conventional form,
\begin{equation}\label{stress}
T_{\mu\nu}=-Pg_{\mu\nu}+(\rho+P)u_{\mu}u_{\nu}~,
\end{equation}
where $\rho$ and $P$ are proper energy density and pressure,
$u_{\mu}$ is 4-velocity with $u^{\mu}u_{\mu}=1$. The energy density
and the pressure of the fluid can be parametrized as \cite{jackiw}
\bea\label{fluid}
\rho&=& f(n)~,\nonumber\\
P&=& nf'(n)-f(n)~, \eea where $f(n)$ is a positive function of $n$.
The introduced variable $n$ can be identified with the number
density and the prime represents derivative with $n$. The equations
of motion are just the covariant conservation equations of the
momentum tensor, $\nabla_{\mu}T^{\mu\nu}=0$. In spatially flat FRW
spacetime
\begin{equation}
   ds^{2}=a^2(\tau)(d\tau^{2} -dx^{i}dx_{i})~~,
\end{equation}
there is only one equation, \be \dot\rho+3\mathcal{H}(\rho+P)=0~,
\ee where the dot is the derivative with the conformal time $\tau$
and $\mathcal{H}=\dot a/a$. Combining Eq.(4) with Eq. (\ref{fluid}),
we get \be f'(n)(\dot n+3\mathcal{H}n)=0~. \ee Since $f'(n)$ does
not vanish everywhere (otherwise it corresponds to the cosmological
constant), one has the conservation equation of the number density
$\dot n+3\mathcal{H}n=0$. In the expanding universe, $n$ will
decrease monotonically with time.

In the following we will demonstrate that the system suffers from
the problem of singularity and classical instability when the
equation of state of the perfect fluid crosses the boundary of $-1$.
Let us assume that the system crosses $-1$ at the point of $n=n_0\neq 0$.
At this point $\rho(n_0)+P(n_0)=0$, $f'(n_0)=0$ and $f'(n)$ will
change the sign after the crossing. So, in the neighborhood of
$n_0$, we can expand $f'(n)$ in terms of $(n-n_0)$. The adiabatic
sound speed square in this neighborhood is \be c_s^2\equiv
\frac{dP}{d\rho}=\frac{nf''(n)}{f'(n)}\simeq\frac{n}{n-n_0}~. \ee We
can see that $c_s^2$ is singular at the crossing point. Moreover, $c_s^2$ is
negative in the region of $n<n_0$. And when $n$ approaches $n_0$
from this side, it will approach $-\infty$. A negative $c_s^2$ will
induce a serious classical instability to the system, the
perturbations on small scales will increase quickly with time and
the late time history of the structure formations will get
significantly modified. This will inevitably lead to the fact that
such models can never be compatible with the observations relevant
to structure formations, such as CMB and LSS.

As shown above it is impossible to realize the quintom scenario
with a single perfect fluid, now we turn to consider the model of scalar field.
The general model of dark energy with a single scalar field and an
arbitrary function of its first derivative
 in the Lagrangian was proposed in Ref. \cite{kessence} and  named
as k-essence. Its Lagrangian usually has a
 non-canonical form
\begin{equation}\label{Skessence}
   \mathcal{L} =P(\phi,X)~~,
\end{equation}
with \begin{equation}\label{Xeq}
   X \equiv  \frac{1}{2} \nabla_{\mu} \phi \nabla^{\mu} \phi~~.
\end{equation}
The energy-momentum tensor of this system has the same form as that
of the single perfect fluid Eq. (\ref{stress}), where
\bea\label{rhokessence}
   & &\rho= 2 X P,_{X} - P~,\nonumber\\
   & &u_{\mu}=\frac{\nabla_{\mu}\phi}{\sqrt{2X}}~.
\eea Let's see under what conditions the system will be able to
cross the barrier of $w=-1$. In order to do that, one requires
$\rho+P$ to vanish at a point of $(\phi_0, X_0)$ and change the sign
after the crossing. This can only be achieved by requiring
$P,_X(\phi_0, X_0)=0$ and $P,_X$ has different signs before and
after the crossing since $X$ cannot be negative. The covariant
conservation law of the energy-momentum tensor gives the equation of
motion, \be
(P,_Xg^{\mu\nu}+P,_{XX}\nabla^{\mu}\phi\nabla^{\nu}\phi)\nabla_{\mu}\nabla_{\nu}\phi+\rho,_{\phi}=0~.
\ee From this equation, we obtain the equation for the background
field \be
\rho,_X[\ddot\phi+(3c^2_{sk}-1)\mathcal{H}\dot\phi]+a^2\rho,_{\phi}=0~,
\ee and the perturbation to the first order (neglecting the metric
perturbations for the time being): \be\label{eqofu} \ddot
u+[-c^2_{sk}\nabla^2-\ddot z/z-3c^2_{sk}(\dot
\mathcal{H}-\mathcal{H}^2)]u=0~, \ee where \be u\equiv
az\frac{\delta\phi}{\dot\phi}~,~~~~~z\equiv
\sqrt{\dot\phi^2|\rho,_X|}~, \ee and the effective sound speed is
given by
\begin{equation}\label{Cs^2kessence}
   c^2_{sk} \equiv \frac{P,_X}{\rho,_X} ~~.
\end{equation}
This sound speed $ c^2_{sk}$ is often used in describing the
perturbations of the scalar fields instead of the isentropic sound
speed, which behave differently due to the intrinsic properties of
the scalar fields \cite{KS84}. For a conventional quintessence or
phantom field, $c^2_{sk}\equiv 1$.
The dispersion
relation from Eq. (\ref{eqofu}) is \be \omega^2=c^2_{sk}k^2-\ddot
z/z-3c^2_{sk}(\dot \mathcal{H}-\mathcal{H}^2)~. \ee One of the
conditions for the stability of k-essence perturbations is that
$c^2_{sk}$ must be positive \cite{kessence}. This requires that $\rho,_X$
has the same behavior as that of $P,_X$, i.e., it must vanish at the
crossing point and change the sign after the crossing.

Similar to the analysis in the case of single fluid, we can see that
$\ddot z/z$ diverges at the point $(\phi_0, X_0)$. This
singularity is unavoidable in the perturbation equation and the
physical quantities describing the fluctuations are not well
defined. Generally, $\ddot z$ does not vanish at the crossing point, hence
there exists a region in which $\omega^2<0$ and the perturbation is unstable.
Furthermore, the canonical momentum defined by the
Lagrangian (\ref{Skessence}) is \be \Pi=\frac{\partial P}{\partial
\dot\phi}=P,_X\frac{\dot\phi}{a^2}~. \ee Its derivative with respect
to $\dot\phi$, \be \frac{\partial
\Pi}{\partial\dot\phi}=\frac{\rho,_X}{a^2}~, \ee vanishes at the
point of crossing. This shows that $\dot\phi$ is not a single valued
function of the momentum $\Pi$ and we cannot get a canonical
Hamiltonian transformed from the Lagrangian unambiguously
\cite{susskind}. The theory cannot be quantized in a canonical way.
Hence we have shown that the conventional k-essence model cannot
give rise to $w$ across $-1$. A different proof is given in Ref. \cite{Vikman04}.

We should stress again that in realistic quintom model buildings one
must consider the aspects of perturbations, where there are often
dangerous instabilities in the conventional case. The concordance
cosmology is based on the precise observations where many of them
are tightly connected to the growth of perturbations and we must
ensure the stability of perturbations. If we start with
parametrizations of the scale factor\cite{barrow} or EOS to
construct quintom models, it can be realized arbitrarily if we do
not consider the stability of perturbations. On the other hand when
we start from scalar fields and use some phenomenological
parametrizations it is in some sense very easy to resemble fluid
behavior in the background evolutions. However the stability of
perturbations must be considered.

\subsection{Some Viable Quintom Models}

As we demonstrated above in the conventional cases with a single
fluid or a k-essence one cannot realize a viable model of quintom,
we need to introduce extra degrees of freedom to realize the
transition of $w$ across $-1$. One of the possibilities is a system
including two fluids with one being $w>-1$ and another $w<-1$.
Specifically, consider a model which consists of two Chaplygin
gases \cite{CGAS} with $P_1=-\lambda_1/\rho_1$ and
$P_2=-\lambda_2/\rho_2$, in which $\lambda_1$ and $\lambda_2$ are
positive constants. If $\rho_1^2>\lambda_1$ and
$\rho_2^2<\lambda_2$, one has $0>w_1>-1$ and $w_2<-1$. This system
will cross the boundary of $-1$ at some time because $\rho_2$ is
always increasing and $\rho_1$ decreasing. The sound speed squares
are $c^2_{si}=-w_i>0$ with $i=1,~2$, hence the system will be free
of the difficulties associated with the singularity and the
classical instability which exist in the model of a single fluid.
Furthermore, the final state of this system will be characterized by
$w=-1$, the universe will approach the de Sitter space in
the far future. In such a scenario there
will be no big rip. In the framework of the field theory, the simple
way to introduce the extra degree of freedom for the quintom model
is the double scalar fields model with one being quintessence-like
and one phantom-like. We should point out that when adding extra degrees of freedom
in the above way, this does not help solve the cosmological constant problem and
nor can it help solve the coincidence problem, since for the component where $w<-1$
it cannot have the property of tracking behavior and has to be fine tuned
\footnote{For the k-essence field where  $w<-1$, it also needs to be fine tuned.}.
The above way of introducing more components
provides the simplest possibility of quintom model building.

There is another possibility of introducing the extra degrees of
freedom for the realization of the transition from the quintessence
phase to the phantom phase. This is the model proposed in Ref.
\cite{MFZ05} by introducing higher derivative operators to the
Lagrangian. Specifically in \cite{MFZ05} we considered a model with
the Lagrangian \be\label{lagrangian} \mathcal{L}=-{1\over
2}\nabla_{\mu}\phi\nabla^{\mu}\phi+{c\over
2M^2}\Box\phi\Box\phi-V(\phi)~, \ee where $\Box\equiv
\nabla_{\mu}\nabla^{\mu}$ is the d'Alembertian operator. The term
related to the d'Alembertian operator is absent in the quintessence,
phantom and the k-essence model, which is the key to make the model
possible for $w$ to cross over $-1$. We have proven in \cite{MFZ05}
 this
Lagrangian is equivalent to an effective two-field
model\be\label{alagrangian} \mathcal{L}= -{1\over
2}\nabla_{\mu}\psi\nabla^{\mu}\psi+{1\over
2}\nabla_{\mu}\chi\nabla^{\mu}\chi -V(\psi-\chi)-{M^2\over
2c}\chi^2~, \ee with the following definition\bea
\chi &=&\frac{c}{M^2}\Box\phi\label{change}~,\\
\psi &=&\phi+\chi~. \eea Note that the redefined fields $\psi$ and
$\chi$ have opposite signs in their kinetic terms. One might be able
to derive the higher derivative terms in the effective Lagrangian
(18) from fundamental theories. In fact it has been shown in the
literature that this type of operators does appear as some quantum
corrections or due to the non-local physics in the string theory
\cite{simon,woodard,gross}. With the higher derivative terms to the
Einstein gravity, the theory is shown to become renormalizable
\cite{stelle} which has attracted many attentions. In fact the
canonical form for the higher derivative theory has been put forward
by Ostrogradski about one and a half century ago
\cite{ostrogradski}. In short, it is interesting and worthwhile to
study further the implications of models with higher derivatives in
cosmology (for a recent study see e.g. \cite{relevnt1}).

\section{Perturbations of the quintom model}
The quintom scenario as we have argued above needs extra degrees of
freedom to the conventional models of a single scalar field, such as
quintessence, phantom and k-essence and the simple realization of
the quintom is a model with two scalar fields or ``equivalently" two
scalar fields for the case with the higher derivative operators.
 In
the discussions below on the perturbations we will restrict
ourselves to the two-field model of quintom with the following
lagrangian:
\begin{equation}\label{quintomlag}
    \mathcal{L}=\mathcal{L}_{Q}+\mathcal{L}_{P}
\end{equation}
 where \begin{equation}\label{qlag}
    \mathcal{L}_{Q}=\frac{1}{2}\partial_{\mu}\phi_{1}\partial^{\mu}\phi_{1}-V_{1}(\phi_{1})
\end{equation}
describes the quintessence component, and
\begin{equation}
\mathcal{L}_{P}=-\frac{1}{2}\partial_{\mu}\phi_{2}\partial^{\mu}\phi_{2}-V_{2}(\phi_{2})
\end{equation}
for the phantom component. The background equations of motion for
the two scalar
 fields $ \phi_i ( i=1, 2) $ are
\begin{equation}\label{phiEOM}
      \ddot{\phi_i} + 2 \mathcal{H} \dot{\phi_i} \pm a^2 \frac{\partial  V_i}{\partial \phi_i}=
      0~~~,
\end{equation}
where the positive sign is for the quintessence and the minus sign
for the phantom. In general there will be couplings between the two
scalar fields, here for simplicity
we neglect them.

For a complete study on the perturbations, the fluctuations of the
fields as well as those of the metric need to be considered. In the
conformal Newtonian gauge the perturbed metric is given by
\begin{equation}\label{lineelecon}
ds^{2}=a^2(\tau)[(1+2\Psi)d\tau^{2} - (1-2\Phi)dx^{i}dx_{i}]~~,
\end{equation}
Using the notations of \cite{ma}, the perturbation equations
satisfied by each of the components of the quintom model (22) are:
\bea
    \dot\delta_{i}&=&-(1+w_{i})(\theta_{i}-3\dot{\Phi})
    -3\mathcal{H}(\frac{\delta
    P_{i}}{\delta\rho_{i}}-w_{i})\delta_{i}~~,\label{dotdelta}\\
\dot\theta_{i}&=&-\mathcal{H}(1-3w_{i})\theta_{i}-\frac{\dot{w_{i}}}{1+w_{i}}\theta_{i}
    +k^{2}(\frac{\delta
    P_{i}/\delta\rho_{i}}{1+w_{i}}\delta_{i}-\sigma_{i} + \Psi)~~,\label{dottheta}
\eea where
\begin{equation}
\theta_{i}=(k^{2}/\dot{\phi_{i}}) \delta\phi_{i}, ~~~\sigma_{i}=0~~,
\end{equation}
\begin{equation}
w_{i}=\frac{P_{i}}{\rho_{i}}~,
\end{equation} and
\begin{equation}\label{prho2}
    \delta P_{i}=\delta\rho_{i}-2V_{i}'\delta\phi_i=\delta\rho_{i}+ \frac{\rho_{i}\theta_{i}}
      {k^{2}}[3\mathcal{H}(1-w_{i}^{2})+\dot{w_{i}}]~~.
\end{equation}

Combining Eqs. (\ref{dotdelta}), (\ref{dottheta}) and (\ref{prho2}),
we have \bea
    \dot\delta_{i}&=&-(1+w_{i})(\theta_{i}-3\dot{\Phi})
    -3\mathcal{H}(1-w_{i})\delta_{i}
    -3\mathcal{H}\frac{\dot w_{i}+3
    \mathcal{H}(1-w_{i}^{2})}{k^{2}}\theta_{i}~,\label{delta2}\\
\dot\theta_{i}&=&2\mathcal{H}\theta_{i}+\frac{k^{2}}{1+w_{i}}\delta_{i}+k^2\Psi~.\label{theta2}
\eea Since the quintom model in (22) is essentially a combination of
a quintessence and a phantom field, one obtains the perturbation
equations of quintom by combining the equations above together. The
corresponding variables for the quintom system are
\begin{equation}\label{wq}
    w_{quintom}=\frac{\sum_i P_{i}}{\sum_i \rho_{i}}~,
\end{equation}
\begin{equation}\label{delta}
   \delta_{quintom}=\frac{\sum_i\rho_{i}\delta_{i}}{\sum_i
   \rho_{i}}~,
\end{equation} and
\begin{equation}\label{theta}
    \theta_{quintom}=\frac{\sum_i(\rho_{i}+p_{i})\theta_{i}}{\sum_i
    (\rho_{i}+P_{i})}~~.
\end{equation}
Note that for the quintessence component, $-1\leq w_{1}\leq 1$,
while for the phantom component, $w_{2}\leq-1$.

With the two fields the quintom model in (22) will be characterized
by the potential $V_i$ in (23) and (24). In this paper we take
$V_i(\phi_i)=\frac{1}{2} m^2_i \phi^2_i$.
 In general the perturbations of $\phi_i$ today stem from
two origins, the adiabatic and the isocurvature modes. If we use the
gauge invariant variable $\zeta_i=-\Phi-\mathcal{H}\frac{\delta
\rho_i}{\dot{\rho_i}}$ instead of $\delta_i$, and the relation
$\Phi=\Psi$ in the universe without anisotropic stress, the
equations (\ref{delta2}) and (\ref{theta2}) can be rewritten as,
\bea
    \dot\zeta_{i}&=&-\frac{\theta_{i}}{3}-C_i(\zeta_i+\Phi+\frac{\mathcal{H}}{k^2}\theta_i)~,\label{mdelta2}\\
\dot\theta_{i}&=&2\mathcal{H}\theta_{i}+k^{2}(3\zeta_i+4\Phi)~,\label{mtheta2}
\eea where \be\label{definec} C_i=\frac{\dot
w_i}{1+w_i}+3\mathcal{H}(1-w_i)=\partial_0[\ln(a^6|\rho_i+p_i|)]~.
\ee $\zeta_{\alpha}$ is the curvature perturbation on the
uniform-density hypersurfaces for the $\alpha$-component in the universe \cite{wands}. Usually, the isocurvature
perturbations of $\phi_i$ are characterized by the differences
between the curvature perturbation of the uniform-$\phi_i$-density
hypersurfaces and that of the uniform-radiation-density
hypersurfaces, \be S_{ir}\equiv 3(\zeta_i-\zeta_r)~, \ee where the
subscript $r$ represents radiation. In this paper, we assume there
is no matter isocurvature perturbations, so $\zeta_m=\zeta_r$.
Eliminating $\zeta_i$ in equations (\ref{mdelta2}) and
(\ref{mtheta2}), we obtain a second order equation for $\theta_i$,
\be\label{2theta} \ddot
\theta_i+(C_i-2\mathcal{H})\dot\theta_i+(C_i\mathcal{H}-2\dot\mathcal{H}+k^2)\theta_i=k^2(4\dot\Phi+C_i\Phi)~.
\ee This is an inhomogeneous differential equation, the general
solution to it is the sum of a general solution to its homogeneous part and a special integration.
In the following, we will show
that the special integration corresponds to the adiabatic
perturbation. Before the era of dark energy domination, the universe
is dominated by some background fluids, for instance, the radiation
or the matter. The perturbation equations of the background fluid
are, \bea
& &\dot\zeta_f=-\theta_f/3~,\nonumber\\
&
&\dot\theta_f=-\mathcal{H}(1-3w_f)\theta_f+k^2[3w_f\zeta_f+(1+3w_f)\Phi]~.
\eea From the Poisson equation \be
-\frac{k^2}{\mathcal{H}^2}\Phi=\frac{9}{2}\sum_{\alpha}\Omega_{\alpha}(1+w_{\alpha})
(\zeta_{\alpha}+\Phi+\frac{\mathcal{H}}{k^2}\theta_{\alpha})\simeq\frac{9}{2}(1+w_{f})
(\zeta_{f}+\Phi+\frac{\mathcal{H}}{k^2}\theta_{f})~, \ee we have
approximately on large scales, \be
\Phi\simeq-\zeta_f-\frac{\mathcal{H}}{k^2}\theta_{f}~. \ee Combining
these equations above with $\mathcal{H}=2/[(1+3w_f)\tau]$, we get
(note numerically $\theta_f\sim \mathcal{O}(k^2)\zeta_f$) \bea
& &\zeta_f=-\frac{5+3w_f}{3(1+w_f)}\Phi={\rm Const.}~,\nonumber\\
& &\theta_f=\frac{k^2 (1+3w_f)}{3(1+w_f)}\Phi\tau~. \eea So, we can
see from Eq. (\ref{2theta}) that there is a special solution to it
which is given approximately on large scales by , \be
\theta_i^{ad}=\theta_f~, \ee and from Eq. (\ref{mtheta2}) we have,
\be \zeta_i^{ad}=\zeta_f~. \ee This shows that the special
integration to Eq. (\ref{2theta}) has the meaning that it
corresponds to the adiabatic perturbation. Hence, for the sake of
isocurvature perturbations of $\phi_i$, we can only consider the
solution to the homogeneous part of Eq. (\ref{2theta}),
\be\label{22theta} \ddot
\theta_i+(C_i-2\mathcal{H})\dot\theta_i+(C_i\mathcal{H}-2\dot\mathcal{H}+k^2)\theta_i=0~.
\ee These solutions are represented by $\theta_i^{iso}$ and
$\zeta_i^{iso}$. The relation between them is \be\label{iso}
\zeta_i^{iso}=\frac{\dot\theta_{i}^{iso}-2\mathcal{H}\theta_{i}^{iso}}{3k^{2}}~.
\ee  Since the general solution of $\zeta_i$ is \be
\zeta_i=\zeta_i^{ad}+\zeta_i^{iso}=\zeta_r+\zeta_i^{iso}~, \ee the
isocurvature perturbations are simply $S_{ir}=3\zeta_i^{iso}$.

In order to solve Eq. (\ref{22theta}), we need to know the forms of
$C_i$ and $\mathcal{H}$ as functions of time $\tau$. For this
purpose, we solve the background equations (\ref{phiEOM}). In
radiation dominated period, $a=A\tau~,~\mathcal{H}=1/\tau$ and we
have \be\label{phi1}
\phi_1=\tau^{-1/2}[A_{1}J_{1/4}(\frac{A}{2}m_1\tau^2)+A_{2}J_{-1/4}(\frac{A}{2}m_1\tau^2)]~,
\ee and \be\label{phi2}
\phi_2=\tau^{-1/2}[\tilde{A}_{1}I_{1/4}(\frac{A}{2}m_2\tau^2)+\tilde{A}_{2}I_{-1/4}(\frac{A}{2}m_2\tau^2)]~,
\ee respectively, where $A$, $A_i$ and $\tilde{A}_i$ are constants,
$J_{\nu}(x)$ is the $\nu$th order of Bessel function and
$I_{\nu}(x)$ is the $\nu$th order of modified Bessel function.
Usually the masses are small in comparison with the expansion rate
in the early universe $m_i\ll \mathcal{H}/a$, we can approximate the
(modified) Bessel functions as $J_{\nu}(x)\sim x^{\nu}(c_1+c_2x^2)$
and $I_{\nu}(x)\sim x^{\nu}(\tilde{c}_1+\tilde{c}_2x^2)$. We note
that $J_{-1/4}$ and $I_{-1/4}$ are divergent when $x\rightarrow 0$.
Given these arguments one can see that this requires large initial
values of $\phi_i$ and $\dot\phi_i$ if $A_2$ and $\tilde{A}_2$ are
not vanished. If we choose small initial values, which is the
natural choice if the dark energy fields are assumed to survive
after inflation, only $A_1$ and $\tilde{A}_1$ modes exist, so
$\dot\phi_i$ will be proportional to $\tau^3$ in the leading order.
Thus, the parameters $C_i$ in equation (\ref{definec}) will be
$C_i=10/\tau$ (we have used $|\rho_i+p_i|=\dot\phi_i^2/a^2$). So, we
get the solution to Eq. (\ref{22theta}), \be
\theta_i^{iso}=\tau^{-4}[D_{i1}\cos(k\tau)+D_{i2}\sin(k\tau)]~. \ee
$\theta_i^{iso}$ oscillates with an amplitude damping with the
expansion of the universe. The isocurvature perturbations
$\zeta_i^{iso}$ decrease rapidly. If we choose large initial values
for $\phi_i$ and $\dot \phi_i$, $A_2$ and $\tilde{A}_2$ modes are
present, $\dot\phi_i$ will be proportional to $\tau^{-2}$ in the
leading order and $C_i=0$. Now the solution to Eq. (\ref{22theta})
is \be \theta_i^{iso}=\tau[D_{i1}\cos(k\tau)+D_{i2}\sin(k\tau)]~.
\ee $\theta_i^{iso}$ will oscillate with a increasing amplitude, so
$\zeta_i^{iso}$ remains constant on large scales.

Similarly, in matter dominated era,
$a=B\tau^2~,~\mathcal{H}=2/\tau$, the solutions for the fields
$\phi_i$ are \be\label{phi11}
\phi_1=\tau^{-3}[B_{1}\sin(\frac{B}{3}m_1\tau^3)+B_{2}\cos(\frac{B}{3}m_1\tau^3)]~,
\ee and \be\label{phi22}
\phi_2=\tau^{-3}[\tilde{B}_{1}\sinh(\frac{B}{3}m_2\tau^3)+\tilde{B}_{2}\cosh(\frac{B}{3}m_2\tau^3)]~,
\ee respectively. We get the same conclusions as those reached by
the above analysis for the radiation dominated era. If we choose
small initial values at the beginning of the matter domination, we
will get the isocurvature perturbations in $\phi_i$  decrease with
time. On the contrary for large initial values, the isocurvature
perturbations remain constant on large scales. This conclusion is
expectable. In the case of large initial velocity, the energy
density in the scalar field is dominated by the kinetic term and it
behaves like the fluid with $w=1$. The isocurvature perturbation in
such a fluid remains constant on large scales. In the opposite case,
however, the energy density in the scalar field will be dominated by
the potential energy due to the slow rolling. It behaves like a
cosmological constant, and there is only tiny isocurvature
perturbation in it.

We have seen that the isocurvature perturbations in
quintessence-like or phantom-like field with quadratical potential
decrease or remain constant on large scales depending on the initial
velocities. In this sense the isocurvature perturbations are stable
on large scales. The amplitude of these perturbations will be
proportional to the value of Hubble rate evaluated during the period
of inflation $H_{inf}$ if their quantum origins are from inflation.
For a large $H_{inf}$ isocurvature dark energy perturbations may be
non-negligible and will contribute to the observed CMB
anisotropy\cite{KMT01,MT04}. In the cases discussed here, however,
these isocurvature perturbations are negligible. Firstly, large
initial velocities are not possible if these fields survive after
inflation as mentioned above. Secondly, even though the initial
velocities are large at the beginning of the radiation domination,
these velocities will be reduced to a small value due to the small
masses and the damping effect of Hubble expansion. In general the
contributions of dark energy isocurvature perturbations are not very
large\cite{GW05} and here for simplicity we assume $H_{inf}$ is
small enough that the isocurvature contributions are
negligible\footnote{We assume in the next section when the mass of
quintessence is larger by an order and oscillates during late time
evolutions, the adiabatic condition still satisfies well.}. Thus we
will concentrate on in next sections the effects of the adiabatic
perturbations of the quintom model with two scalars considered in
this paper.

\section{Signatures of quintom and the effects of perturbations on observations}

Based on the perturbation equations(\ref{delta})
 and (\ref{theta}), we modify the code
of CAMB \cite{camb} and will study preliminarily in this chapter the observational
signatures of quintom. Throughout this paper we assume a flat
universe. In showing the illustrative effects for quintom we have
assumed the fiducial background parameters to be $\Omega_{b}=0.042,
\Omega_{c}=0.231, \Omega_{DE}=0.727$, where $DE$ denotes dark energy
and today's Hubble constant is fixed at $H_{0}=69.255$ km/s
Mpc$^{-2}$. We will calculate the effects of perturbed
quintom on CMB and LSS.

\begin{figure}[htbp]
\begin{center}
\includegraphics[scale=0.9]{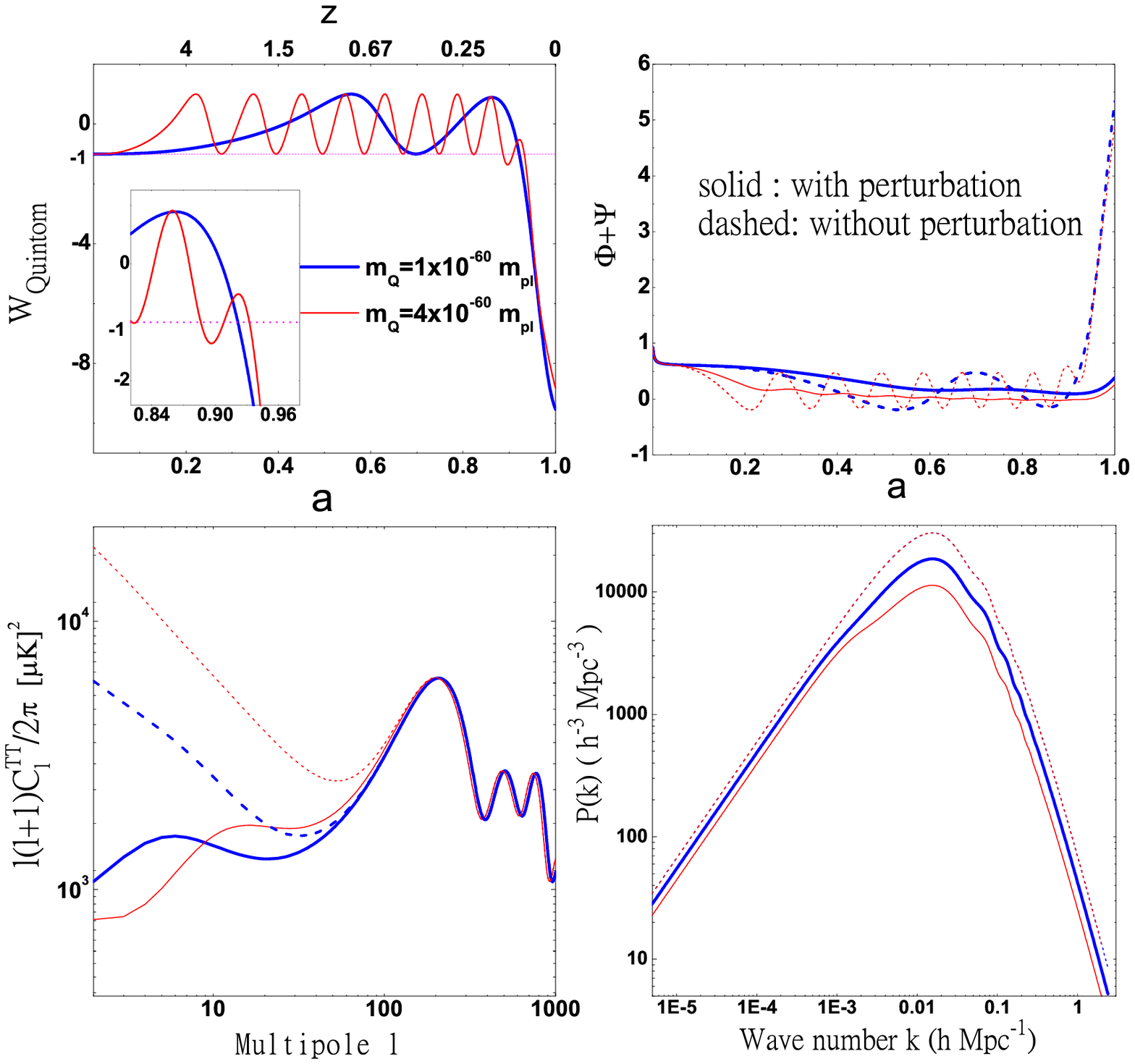}
\caption{Effects of the two-field oscillating quintom on the
observables. The mass of the phantom field is fixed at
$2.0\times10^{-60} m_{pl}$ and the mass of the quintessence field
are $10^{-60} m_{pl}$(thicker line) and $4.0\times10^{-60}
m_{pl}$(thinner line) respectively. The upper right panel
illustrates the evolution of the metric perturbations $\Phi+\Psi$ of
the two models when with(solid lines) and without(dashed lines) dark
energy perturbations. The scale is $k\sim10^{-3}$ Mpc$^{-1}$. The
lower left panel shows the CMB effects and the lower right panel
delineates the effects on the matter power spectrum when with(solid
lines) and without(dashed lines) dark energy
perturbations.}\label{Fig1}
\end{center}
\end{figure}

In the quintom model we focus on there are two
parameters: one is the quintessence mass and the other being the phantom
mass. When the mass of quintessence is heavier than Hubble parameter
the field will start to oscillate and consequently one will get an
oscillating quintom. In the numerical discussions we will fix the
mass of the phantom field to be $m_P\sim 2.0 \times 10^{-60} m_{pl}$.
We vary the quintessence mass with the typical values being $m_Q=10^{-60} m_{pl}$
and $4 \times 10^{-60} m_{pl}$
respectively. We plot in Fig. 1 the equations of state as function
of the scale factor for the above two sets of the parameters and their corresponding
effects on the observations.
 One can see the obvious
oscillating feature of quintom as the mass of quintessence
component goes heavier. After touching the $w=-1$ pivot for
several times, $w$ crosses $-1$ consequently where the phantom part
dominates dark energy. The quintom field modifies the metric perturbations:
$\delta g_{00}=2a^{2}\Psi,\delta g_{ii}=2a^{2}\Phi \delta_{ij}$ and consequently contribute to the
late time Integrated Sachs-Wolfe (ISW) effect. The ISW effect is an
integrant of $\dot{\Phi}+\dot{\Psi}$ over conformal time and
wavenumber k. The above two quintom models yield quite different
evolving $\Phi+\Psi$ as shown in the right panel of Fig.\ref{Fig1},
where the scale is $k\sim 10^{-3}$ Mpc$^{-1}$. We can see the late
time ISW effects differ significantly when dark energy perturbations
are taken into account(solid lines) or not(dashed lines).

ISW effects take an important part on large angular scales of CMB
and on the matter power spectrum of LSS. For a constant EOS of
phantom Ref.\cite{WL03} has shown that the low multipoles of CMB
will get significantly enhanced when dark energy perturbations are neglected.
On the other hand for a matter like scalar field where the equation of state is around zero,
perturbations will also play an important role on the large scales
of CMB, as shown in ref.\cite{CDS98}. Our results on CMB and LSS
reflect the two combined effects of phantom and oscillating
quintessence. Note that in the early studies of quintessence effects
on CMB, one usually considers a constant $w_{eff}$
instead:
\begin{equation}\label{weff}
    w_{eff}\equiv\frac{\int da \Omega(a) w(a)}{\int da \Omega(a)}~~,
\end{equation}
however this is not enough for the study of effects on SN, nor for
CMB when the EOS of dark energy has a very large variation with
redshift, such as the model of oscillating quintom considered in this paper.

\begin{figure}[htbp]
\begin{center}
\includegraphics[scale=0.9]{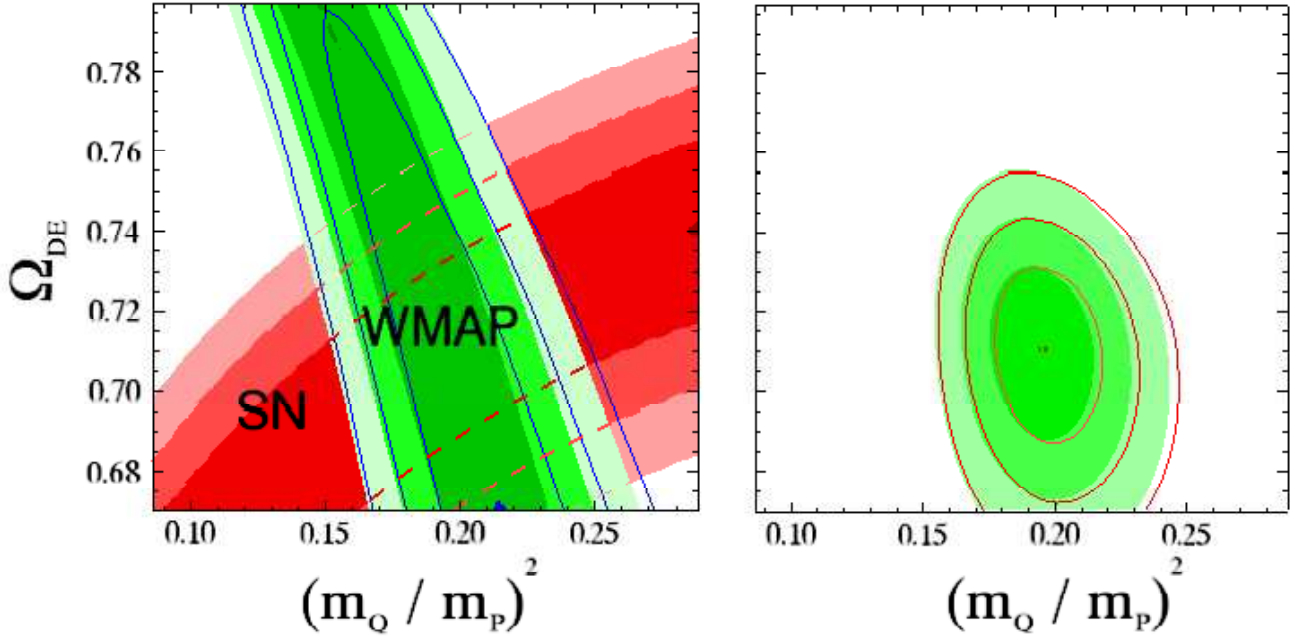}
\vspace{-11.5cm} \caption{3$\sigma$ WMAP and SN constraints on
two-field quintom model shown together with the best fit values.
$m_Q$ and $m_P$ stand for the mass of quintessence and phantom
respectively. We have fixed $m_P\sim 6.2 \times 10^{-61} m_{pl}$ and
varied the value of $m_Q$. Left panel: Separate WMAP and SN
constraints. The green(shaded) area is WMAP constraints on models
where dark energy perturbations have been included and the blue
area(contour with solid lines) is without dark energy perturbations.
Right panel: Combined WMAP and SN constraints on the two-field
quintom model with perturbations(green/shaded region) and without
perturbations(red region/contour with solid lines). }\label{Fig2}
\end{center}
\end{figure}

To face the oscillating model of quintom with the current observations,
we make a preliminary fitting to the first year Wilkinson
Microwave Anisotropy Probe(WMAP) TT and the TE
temperature--polarization cross-power spectrum as well as the recently
released 157 ``Gold" SN data\cite{Riess04}. Following
Refs.\cite{smalll,Fengl1} in all the fittings below we will fix
$\tau=0.17$, $\Omega_m h^2=0.135$ and $\Omega_b h^2=0.022$, we
set the spectral index as $n_S=0.95$ and the amplitude of the
primordial spectrum will be used as a continuous parameter. In the
fittings of oscillating quintom we've fixed the mass of phantom to
be $m_P\sim 6.2 \times 10^{-61} m_{pl}$. Fig.\ref{Fig2} delineates
3$\sigma$ WMAP and SN constraints on the two-field quintom model, it also
shows the corresponding best fit values. In the labels $m_Q$ and
$m_P$ stand for the mass of quintessence and phantom respectively.
The left panel of Fig.\ref{Fig2} shows the separate WMAP and SN
constraints. The green(shaded) area is WMAP constraints on models
where dark energy perturbations have been included and the blue
area(contour with solid lines) is without dark energy perturbations.
The perturbations of dark energy have no effects on the geometric
constraint of SN. The right panel shows the combined WMAP and SN
constraints on the two-field quintom model with perturbations
(green/shaded region) and without perturbations (red region/contour
with solid lines). We find the confidence regions do show a large
difference when the perturbations of dark energy have been taken
into account or not.

\begin{figure}[htbp]
\begin{center}
\includegraphics[scale=0.9]{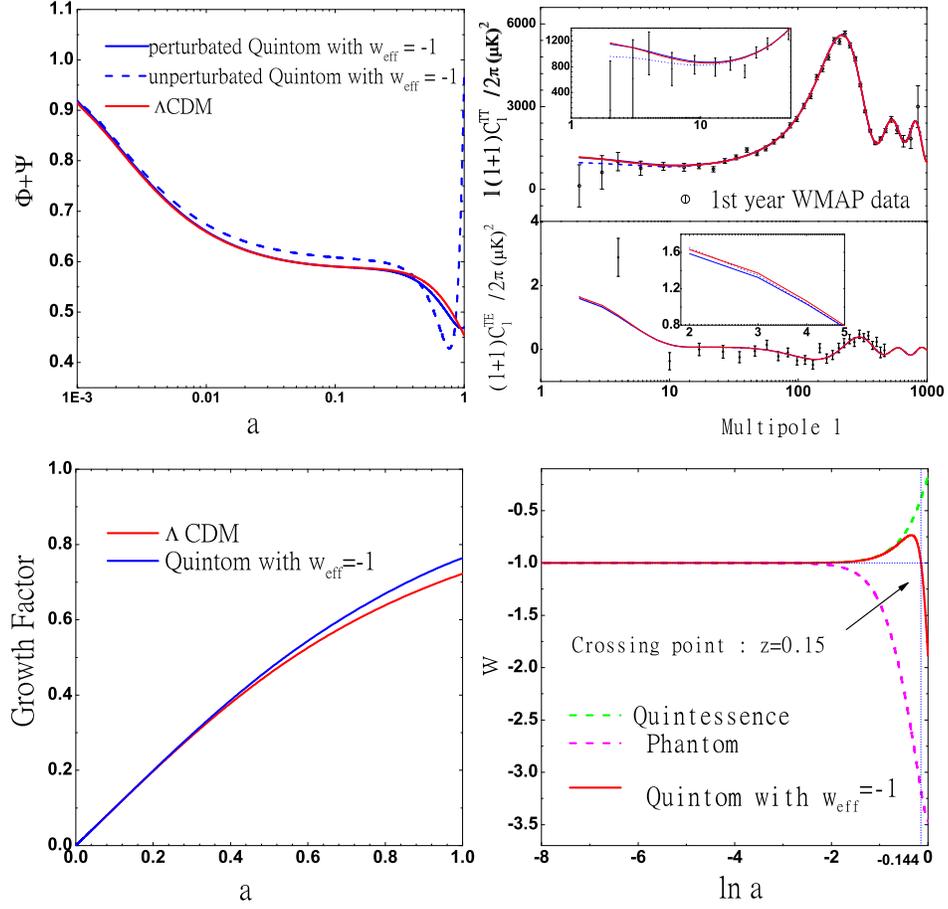}
\caption{Effects of the two-field quintom model where $w_{eff}=-1$
compared with cosmological constant in CMB(WMAP), the metric
perturbations $\Phi+\Psi$(the scale is $k\sim10^{-3}$ Mpc$^{-1}$)
and the linear growth factor. The binned error bars in the upper
right panel are WMAP TT and TE data \cite{Kogut03}.}\label{Fig3}
\end{center}
\end{figure}

So far we have investigated the imprints of oscillating quintom  on
CMB and LSS. Now we consider another example where $w$ crosses $-1$
smoothly without oscillation. It is interesting to study the effects
of this type of  quintom model with its effective equation of state
defined in (57) exactly equal to $-1$ on CMB and matter power
spectrum. This study will help to distinguish the quintom model from
the cosmological constant. We have realized such a model of quintom in the
lower right panel of Fig.\ref{Fig3}, which can be easily given in
the two-field model with lighter quintessence mass. In this example
we have set $m_Q\sim 2.6 \times 10^{-61} m_{pl}, m_P\sim 6.2 \times
10^{-61} m_{pl}$. We  assume there are no initial kinetic energy. The initial
values of the quintessence component is set as $\phi_{1i}=0.226 m_{pl}$ and the
phantom part: $\phi_{2i}= 6.64 \times 10^{-3} m_{pl}$. We find the
EOS of quintom crosses $-1$ at $z\sim 0.15$, which is consistent with
the latest SN results.

 The model of quintom, which is mainly favored
by current SN only, needs to be confronted with other observations
in the framework of concordance cosmology. SN making the only direct
detection of dark energy, this model is most promising to be
distinguished from the cosmological constant and other dynamical dark energy
models which do not get across $-1$ by future SN projects
on the low redshift(for illustrations see e.g. \cite{cooray}). This is also the case for the model of quintom in the full parameter space: it can be most directly tested in low
redshift Type Ia supernova surveys. In the upper left panel of
Fig.\ref{Fig3} we delineate the different ISW effects among the
cosmological constant (red/light solid), the quintom model which
gives $w_{eff}=-1$ with (blue/dark solid) and without(blue dashed)
perturbations. Similar to the previous oscillating case, the
difference is very large when switching off quintom perturbations
and much smaller when including the perturbations. In the
upper right panel we find the quintom model cannot be distinguished
from a cosmological constant in light of WMAP. The two models almost
give exactly the same results in CMB TT and TE power spectra when
including the perturbations. We find the difference in CMB is hardly
distinguishable even by cosmic variance.

Given the fact aobve that from CMB observations quintom with $w_{eff} = -1$ makes no
distinctive signatures, now we discuss briefly the signatures in some other observations. To do that
we need to consider the physical observables
which can be affected by the evolving $w$ sensitively. In comparison
with the cosmological constant such a quintom model gives a
different evolution of expansion history of universe, such as altering the epoch of matter-radiation equality. The Hubble expansion
rate $H$ is :
\begin{equation}\label{H}
    H \equiv \frac{\dot{a}}{a^2}=H_{0}[\Omega_{m}a^{-3}+\Omega_{r}a^{-4}+X]^{1/2}
\end{equation}
where X, the energy density ratio of dark energy between the early
epochs and today, is quite different for the $quintom$-CDM and
$\Lambda$CDM. In the $\Lambda$CDM scenario, X is simply a constant
while in general for dark energy models with varying energy density
or EOS,
\begin{equation}\label{omegaquitom}
    X=\Omega_{DE} a^{-3}e^{-3\int w(a)d \ln a} ~~.
\end{equation}
The two models will give different Hubble expansion rates. This is
also the case between the quintom model with $w_{eff} = -1$ in the
left panel of Fig.\ref{Fig3} and a cosmological constant. Different
$H$ leads directly to different behaviors of the growth factor. In the linear
perturbation theory all Fourier modes of the matter density
perturbations grow at the same rate. The matter density
perturbations are independent of $k$:
\begin{equation}\label{detamk}
    \ddot{\delta}_k + \mathcal{H} \dot{\delta}_k - 4 \pi G a^2 \rho_{\rm
M} \delta_k=0 ~~.
\end{equation}
The growth factor $D_{1}(a)$ characterizes the growth of the matter
density perturbations: $D_{1}(a)= \delta_k (a)/\delta_k (a=1)$ and
is normalized to unity today. In the matter-dominated epoch we have
$D_1 (a)=a$. Analytically $D_{1}(a)$ is often approximated by the
Meszaros equation \cite{Dodsbk}:
\begin{equation}\label{d1}
    D_{1}(a)=\frac{5\Omega_{m}H(a)}{2H_{0}}\int^{a}_{0}\frac{da'}{(a'H(a')/H_{0})^{3}}~,
\end{equation}
where we can easily see the difference between the model of quintom
and a cosmological constant due to the different Hubble expansion rates.
More strictly one needs to solve Eq.(\ref{detamk}) numerically. In
the lower left panel of Fig.\ref{Fig3} we show the difference of
$D_1 (a)$ between the quintom with $w_{eff} = -1$ and the cosmological
constant. The difference in the linear growth function is
considerably large in the late time evolution and possibly
distinguishable in future LSS surveys and in weak gravitational
lensing (WGL) observations. WGL has emerged with a direct mapping of
cosmic structures and it has been recently shown that the method of
cosmic magnification tomography can be extremely
efficient\cite{lensmagn}, which leaves a promising future for
breaking the degeneracy between quintom and a cosmological constant.

\section{Perturbations of parametrized quintom and
the effects on the observations}

There have been many studies in the literature in the fittings of
the dark energy with parametrized EOS, such as the linear
parametrization $w=w_0+w_1 z$ \cite{linearP} to SN and other
observations such as CMB and LSS. For the latter observations the
perturbations of dark energy need to be considered.
 However, at the point of
$w=-1$, as pointed out in Section II one would be encountered with
the singularity of the isentropic sound speed. Moreover in the
perturbation equation (\ref{dottheta}) one will get infinite
$\dot{\theta}$. For the physical quantity $(\rho+P)\theta$ in the
model of the single field of quintessence, it is not divergent at
$w=-1$, {\it i.e.} $\theta\rightarrow \infty$ but $ (\rho+P)\theta =
k^2 \dot{\phi} \delta \phi =0$, however for the model with
parametrized EOS one will generically have an unphysical divergence
when $\dot{w}\neq 0$ at the cosmological constant boundary.
The detailed explanation is given as
follows: firstly from Eq.(\ref{dottheta}) one will get infinite
$\dot{\theta}$ and the physical continuity implies that one will
also get $\theta\rightarrow \infty$ at $w=-1$. Introducing the new
physical quantity which is relevant to the CMB observations:
\begin{equation}\label{BFNju}
  \mathcal{V}\equiv (1+w)\theta~~,
\end{equation}
Eqs.(\ref{delta2}, \ref{theta2}) can be rewritten now as:
 \bea
    \dot\delta &=&- \mathcal{V}  + (1+w )3\dot{\Phi}
    -3\mathcal{H}(1-w )\delta
    -3\mathcal{H}\frac{\dot w  / (1+w )+3
    \mathcal{H}(1-w )}{k^{2}} \mathcal{V} ~,\label{delta2nju}\\
\dot \mathcal{V} &=&2\mathcal{H} \mathcal{V} + k^{2}\delta
+\frac{\dot{w }}{1+w } \mathcal{V} +k^2(1+w)\Psi~.\label{theta2nju}
\eea We can easily see that $\dot \mathcal{V}\rightarrow \infty$
when $\dot{w}\neq 0$ at $w=-1$.

We should point out that both the scalar fields and fluids obey the
same form of equations on the evolution of perturbations:
Eqs.(\ref{dotdelta},\ref{dottheta}), and the only difference comes
from the term of $\delta P_i/\delta \rho_i$. If one starts from
Eqs.(\ref{delta2}, \ref{theta2}) and study the effects of dark
energy by parametrizing the EOS, this is equivalent to the
description of the effects of the scalar field and is identical to
work starting with dark energy potentials. If in models with
the parametrized EOS we have $w$ always in the range $[-1,1]$, or
$w\le -1$ for $0<a<\infty$, there will be no unphysical divergence
and this equivalently describes the single field of quintessence or
phantom\cite{guo05}. For example in model with $w=w_0+w_1 \sin(\ln
a)$, if we restrict $w_0=0$ and $|w_1|\le 1$ then Eqs.(\ref{delta2nju},
\ref{theta2nju}) will always be continuous. However when the
parameter space is enlarged to include $\dot{w}\neq 0$ at $w=-1$
Eqs.(\ref{delta2nju}, \ref{theta2nju}) will be unphysical.

We emphasize that the above discussions are valid only for
models with a single field. For models with multi fields we have
shown explicitly in the previous sections the perturbation equations
Eqs.(\ref{delta2}, \ref{theta2}) are \textbf{continuous} during the crossing
of the cosmological constant boundary. It is similar
for models with two fluids or models with two components of
parametrized EOS: $w=\Sigma \Omega_i w_i$ where each component $w_i$
does not evolve across $-1$. This implies, however, in the fitting of
the models to the observational data the parameters introduced for
the EOS should be doubled if allowing the EOS $w$ to vary and get
across $-1$. Certainly this is not practically applicable. It would
be nice to develop a technique to include the perturbations which
approximates well to the quintom, meanwhile not introducing the
extra degrees of freedom to the models considered widely in the
literature with parametrized EOS. We will make a proposal for it below.

First of all we consider a system of quintom with two fields as
above, $\phi_1$ being quintessence-like and $\phi_2$ being
phantom-like, but restrict the EOS of the system not to cross over
$-1$. In this case we will show the background of this system is
equivalent to a model with an effective single scalar field denoted
by $\chi$. By definition the pressure P and energy density $\rho$ of
the $\chi$ field should be equal to the two-field case. When the
kinetic term of $\phi_1$ is larger than that of the phantom part
$\phi_2$, the whole system of dark energy gives rise to an EOS
larger than $-1$ and the effective $\chi$ behaves like a quintessence.
On the contrary when $\dot{\phi_1}^2-\dot{\phi_2}^2\le 0$, $\chi$ is
a phantom field. Hence the kinetic and potential terms of $\chi$, in
terms of $\phi_1$ and $\phi_2$, can be expressed as
\begin{equation}\label{equvkQ}
 \pm \dot{\chi}^2 = \dot{\phi_1}^2 - \dot{\phi_2}^2 ~~
\end{equation}
and
\begin{equation}\label{equvPQ}
 V(\chi)= V(\phi_1) + V(\phi_2)  ~~,
\end{equation}
where the "+" sign in Eq. (\ref{equvkQ}) is for the case where the
total EOS of dark energy is quintessence-like and the "-" sign for
phantom-like evolutions. We can directly reconstruct the potential
and time evolutions of $\chi$. For example if we set the potentials
of the two fields to be both linear:
\begin{equation}\label{linearVf}
  V_i(\phi_i) = V_{0i} + \lambda_i \phi_i ~~,
\end{equation}
in the early epochs of radiation and matter domination dark energy
fields are slow rolling and
\begin{equation}\label{dotsqp}
{\phi_1}'\sim - \lambda_1/ 3 H~~, ~~{\phi_2}'\sim  \lambda_2/ 3 H~~,
\end{equation}
where prime denotes derivative respects to the physical time. For
the whole system in the quintessence phase
$\phi_1$ will have a larger kinetic energy, and in the radiation dominant
epoch
\begin{equation}\label{RD}
H=1/2t~~, ~~\chi'= \pm \frac{2}{3}t\sqrt{\lambda^2_1-\lambda^2_2}~~.
\end{equation}
On the other hand from Eq.(\ref{equvPQ}) we have
\begin{equation}\label{equvpPQ}
 V_{,\chi}(\chi) \chi'= V_{,\phi_1}(\phi_1)\phi_1' + V_{,\phi_2}(\phi_2)\phi_2'  ~~,
\end{equation}
combining Eqs.(\ref{dotsqp},\ref{RD},\ref{equvpPQ}) we can easily
get
\begin{equation}\label{equvpPQ1}
 V_{,\chi}(\chi) = \mp \sqrt{\lambda^2_1-\lambda^2_2}  ~~,
\end{equation}
consequently the effective potential of $\chi$ analytically is
\begin{equation}\label{linearVQ}
 V(\chi)= \pm \sqrt{\lambda_1^2 - \lambda_2^2}(\chi-\chi_0)~~,
\end{equation}
where $\chi_0$ can be easily set by the initial conditions of
$\phi_i$ and the sign of ``$+
$" or ``$-$" is somewhat optional. The
arguments above applies for the case when the total EOS of the
system is restricted to be no larger than $-1$, the effective scalar
will behave like phantom.

On the
evolution of perturbations we can see from
Eqs.(\ref{dotdelta},\ref{dottheta}) that the phantom and the quintessence
fields obey the same equations. As shown in Section III although
generically the two field model  would have non-vanishing
isocurvature perturbations we can choose suitable initial conditions
so that the isocurvature contributions can be safely negligible. In
this sense when the total EOS does not evolve across minus unity,
the whole system can be equally described by an adiabatic field:
both the background evolution and the adiabatic perturbations, as
shown similarly in Refs.\cite{Gordon01,Fengl1,Fengl2} in the
inflationary universe.

\begin{figure}[htbp]
\begin{center}
\includegraphics[scale=0.9]{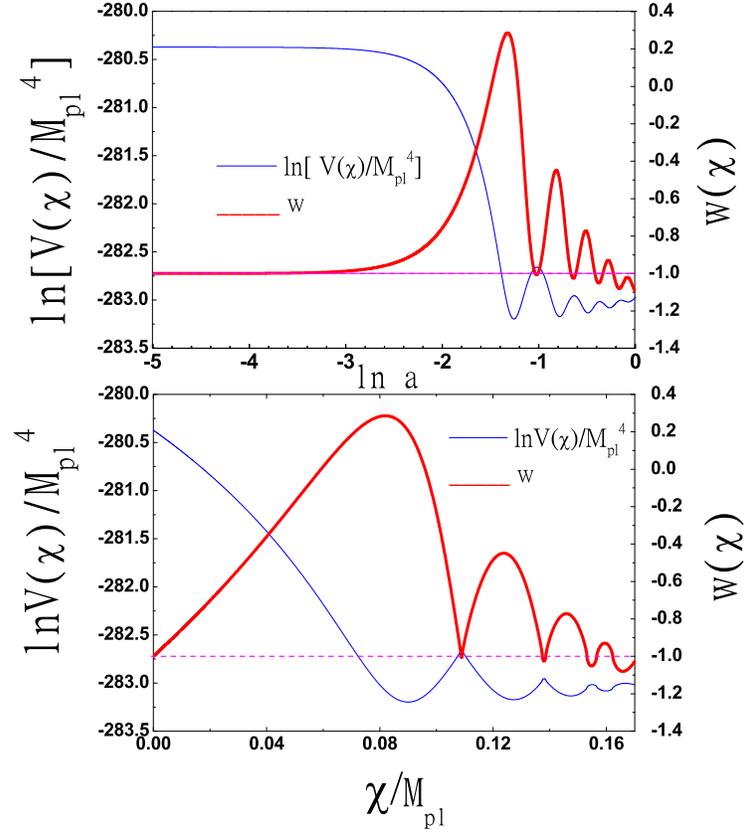}
\caption{ Reconstructions of the effective adiabatic single field
$\chi$ in the framwork of oscillating quintom. The background
parameters have been chosen as $m_{\phi 1}= 2\times 10^{-60} m_{pl},
m_{\phi 2}=  10^{-61} m_{pl}$, initial values are $\phi_{1i}=0.09
m_{pl}, \phi_{2i}=0.45 m_{pl}$ and $\dot{\phi_{1i}}=
\dot{\phi_{2i}}=0 $ early in radiation domination epoch and for this
example we have $\Omega_{\phi 1}=0.2, \Omega_{\phi 2}\sim 0.54,
h\sim 0.68$. The red lines are the total EOS of dark energy and the
blue lines are the total potential. The dashed lines show the
cosmological constant boundary. The upper panel delineates the late
evolutions of the EOS and potential of dark energy and the lower
panel shows the reconstructed values of $\chi$ and its potential,
$\chi$ is a quintessence/phantom scalar when $w$ is above/below the
dashed line. See the text for details.}\label{Fig5}
\end{center}
\end{figure}

We have demonstrated in the previous paragraphs
the equivalence between the two-field quintom model
and the
single scalar field model when the EOS of the system does not cross over
$-1$. However
if
the total equation of state for the double fields does cross over $-1$, this
system will not be able to be described effectively by
a single scalar field. To study the perturbations of the dark energy models with EOS across
$-1$, we introduce a small positive constant $c$ to divide the whole
region of the allowed value of the EOS $w$ into three parts: 1) $ w
> -1 + c$; 2) $-1 + c > w > -1 - c$; and 3) $w < -1 -c $. For 1) the
EOS is always larger than $-1$ and for 3) $w$ is always less than
$-1$. For both cases the system with two fields as shown above can be
described effectively by a single scalar field with a potential
satisfying \cite{DCS02}
\begin{equation}\label{vpp}
\pm a^2
\frac{d^2V}{d\chi^2}=-\frac{3}{2}(1-w)\left[\frac{\ddot{a}}{a}
-\mathcal{H}^2\left(\frac{7}{2}+\frac{3}{2}w\right)\right]
\nonumber \\
+\frac{1}{1+w}\left[\frac{{\dot{w}}^2}{4(1+w)}-
\frac{\ddot{w}}{2}+\dot{w}\mathcal{H}(3w+2)\right]~~,\nonumber\\
\end{equation}where ``$+$" is for the case 1) and ``$-$" for the
case 3). One can see that $\frac{d^2V}{d\chi^2}$ is divergent and
there would be a discontinuity in the derivative $V'$ at the turning
point of $w=-1$, which corresponds to $c \rightarrow 0 $. As an
example in Fig. \ref{Fig5} we give the reconstructed potential of
the effective $\chi$ field for an oscillating quintom. One can see
$\chi$ behaves like quintessence when $w > -1$ and like phantom when $w<
-1$. The reconstructed potential is well defined except in region 2)
when the EOS gets across $-1$, where there is a sharp discontinuity on
$V'(\chi)$.

For the case 2), different from those in 1) and 3), the perturbations
cannot be fully described by a single adiabatic field. However as
we learn from the above, for the realistic quintom models the perturbations in
the region 2) will be continuous and not divergent, {\it i.e.}
$\delta$ and $\theta$ are continuous, and the derivatives of
$\delta$ and $\theta$ are finite. A good approximation to the
perturbation in region 2) is requiring it to match to the regions 1)
and 3) at the boundary. Practically we take $\delta$ and $\theta$ to
be constant matching to regions 1) and 3) at the boundary and set
\begin{equation}\label{dotx}
  \dot{\delta}=0 ~~,~~\dot{\theta}=0 .
\end{equation}
In the numerical calculation the constant $c$ is a very small
number, the approximation above lies in a very close
neighborhood of $w=-1$.
In practice in our numerical calculations we've
limited the range to be $|\Delta w = c |<10^{-5}$.
 Since the region 2)
is \textbf{extremely limited}, neglecting the evolutions of
perturbations as shown in (74) is quite safe and well
approximated. Thus we can use Eqs.(\ref{delta2}, \ref{theta2}) to
study the effects of perturbations in models with parametrized EOS.
We have also numerically checked the validity of Eq.(74) and found
their contributions to the observed CMB and LSS power spectra are
very small. The procedure of our checking is listed as follows:

\begin{enumerate}

\item  Start
with the two-field model of quintom and record $w(a)$, compute CMB
and LSS spectra with perturbations.

\item Build a code in CAMB\cite{camb} to include dark energy
perturbations with
parametrized EOS. Include perturbations by setting
Eq.(\ref{dotx}) and treating $\delta, \theta$ as continuous.

\item
Interpolate w(a) in the code with parametrized EOS, compute CMB and
LSS spectra and make comparisons with the results from step 1.
\end{enumerate}
With this procedure we have considered a model of oscillating
quintom and found the difference is no more than $10^{-4}$, which is
safely negligible.

\begin{figure}[htbp]
\begin{center}
\includegraphics[scale=0.9]{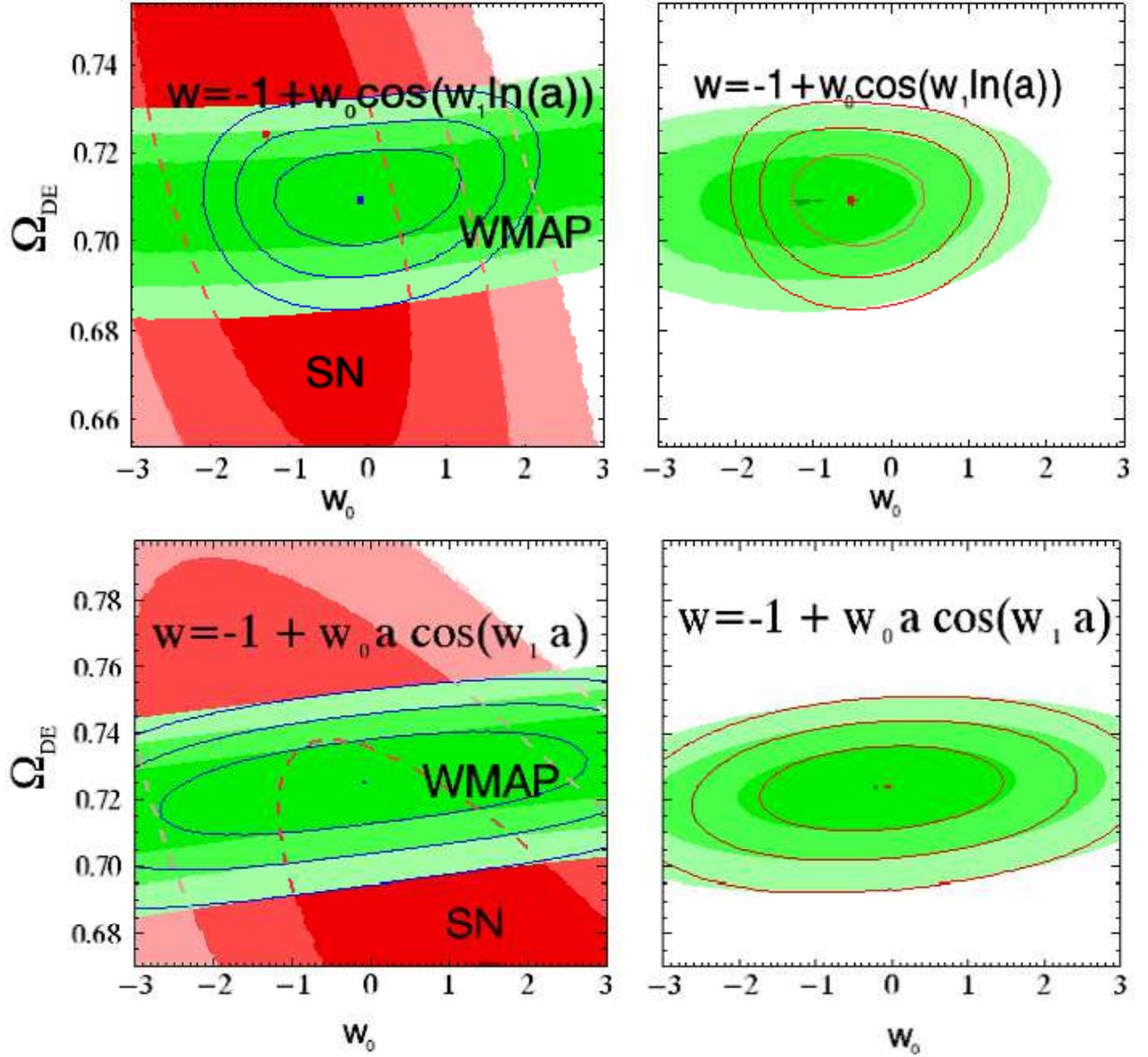}
\vspace{-9cm} \caption{ 3$\sigma$ WMAP and SN constraints on the
parametrized quintom models $w= -1 + w_0 \cos(w_1 \ln a)$ and $w=
-1 + w_0 a \cos(w_1 a)$, shown together with the best fit values. On
the left panels the green(shaded) areas are WMAP constraints on
models where dark energy perturbations have been included and the
blue areas(contours with solid lines) are without dark energy
perturbations. On the right panels models with perturbations are
delineated in green(shaded) regions and the red regions(contour with
solid lines) without perturbations. For illustrations we have fixed
$w_1=20$ in the upper panels and $w_1=50$ in the lower panels.
}\label{Fig6}
\end{center}
\end{figure}

As examples now we study the effects of perturbations for several models with
parametrized EOS in light of WMAP and SN data. The first example is
given by Ref.\cite{FLPZ}, where $w$ is parametrized by
\begin{equation}\label{osceq} w(\ln a)=w_a + w_0 \cos[w_1 \ln
(a/a_c)]
\end{equation}
with $a$ being the scale factor.
This model has a nice feature of  unifying the early inflation and
the current
accelerated expansions.
 In Ref.\cite{FLPZ} the period of
oscillation has been set as long as $\sim 200$ e-folds. It is
interesting to study the consequences with a shorter period. Here
for illustrations we fix $w_a=-1$, $w_1=20$ and $a_c=1$. In the
upper panels of Fig.\ref{Fig6} we show the illustrative fittings
when with and without the perturbations. We can see the parameter
space has been enlarged a lot when including the contributions of
the perturbations. For a second example we parametrize $w$ as
\begin{equation}\label{osceq1} w(\ln a)=w_a + w_0 a \cos[w_1 a
+a_c].
\end{equation}
In the numerical calculation we've fixed  $w_a=-1$, $w_1=50$ and
$a_c=0$. In the lower panels of Fig.\ref{Fig6} we can see the
effects are still very prominent both in the separate and combined
constraints, although not as strong as the example in (75). For the
third example we take $w$ to be non-oscillatory:
\begin{equation}\label{crosseq} w= w_0/(1-\ln a)~~,
\end{equation}
where the original form was firstly proposed in Ref.\cite{CW04}. We
find in our case SN constraints are very weak due to the fixed
background parameters, the 1$\sigma$ regions have not been affected
much by the perturbations, but the 2, 3$\sigma$ regions have been
enlarged significantly when the perturbations are taken into
account.

\section{Conclusions}

In this paper we have studied the perturbations of the dynamical
quintom model of dark energy in a self-consistent way. It is physically significant for the inclusion of quintom perturbations, both on the
theoretical grounds of model buildings and on the fittings to the observations.
Due to the singularities and
instabilities of perturbations at the cosmological constant
boundary, we have shown a new method regarding the impossibility of k-essence as
a viable quintom model. In the realistic quintom model buildings one must include the
perturbations. In general one needs to add extra degrees of freedom to
realize the model of quintom. In the two-field model and the model
with a d'Alembertian operator the isocurvature contributions may be
safely negligible in the simplest case. We have considered the
implications of quintom perturbations on the observations of CMB and LSS.
We have shown that the parameter space is different when one includes the perturbations of
dark energy or not. In trying to constrain dark energy in a model
independent way we have also proposed a method to include the
perturbations for models of dark energy with parametrized EOS
across $-1$. With some specific examples of the parametrized EOS,
we show that the parameter space which characterizes
the properties of the model will get enlarged in general when
including the perturbations. This will lead to important consequences
in the phenomenological studies on the cosmological imprints of dynamical dark energy, including the model of quintom. A thorough investigation of current
constraints on the quintom model of dark energy where dark
perturbations are taken into account is beyond the scope of current
paper and will be presented elsewhere \cite{toappear,zhang05}.

Overall, a dynamical quintom model is favored by current SN data and
not ruled out by the combined observational constraints. There are
still some inconsistencies today among different observations in the
precision cosmology and the concordance $\Lambda$CDM model has not
yet fitted well to the observations in a high enough confidence
level, in this sense we might not be adopting the Ockham's razor
with a cosmological constant. When we start from a $\Lambda$CDM
model in the probe of our universe we cannot achieve more subtle
physics beyond that. This is necessary to bear in mind for us to
understand the nature of dark energy with the accumulation of the
observational data.

{\bf Acknowledgements:} We thank the anonymous referee for helpful suggestions.
 We thank Robert Brandenberger, Xue-Lei Chen,
Zuhui Fan,  Pei-Hong Gu, Hong Li, Hiranya Peiris, Yunsong Piao,
Yong-Zhong Xu and Peng-Jie Zhang for helpful discussions. We
acknowledge the using of CMBfast\cite{cmbfast,IEcmbfast} in our
early studies and CAMB\cite{camb,IEcamb} for all the numerical
calculations. In the fitting to WMAP we've used the code developed
in Ref. \cite{Verde03}. We thank Antony Lewis for early
miscellaneous help and discussions on the cosmocoffee\cite{coffee}.
This work is supported in part by National Natural Science
Foundation of China under Grant Nos. 90303004 and 19925523 and by
Ministry of Science and Technology of China under Grant No. NKBRSF
G19990754. B. F. would like to thank the hospitality of the National
Astronomical Observatories, Chinese Academy of Sciences where part
of this work was finished and M. L. is supported by Alexander von
Humboldt Foundation.

\begin{figure}[htbp]
\begin{center}
\includegraphics[scale=0.9]{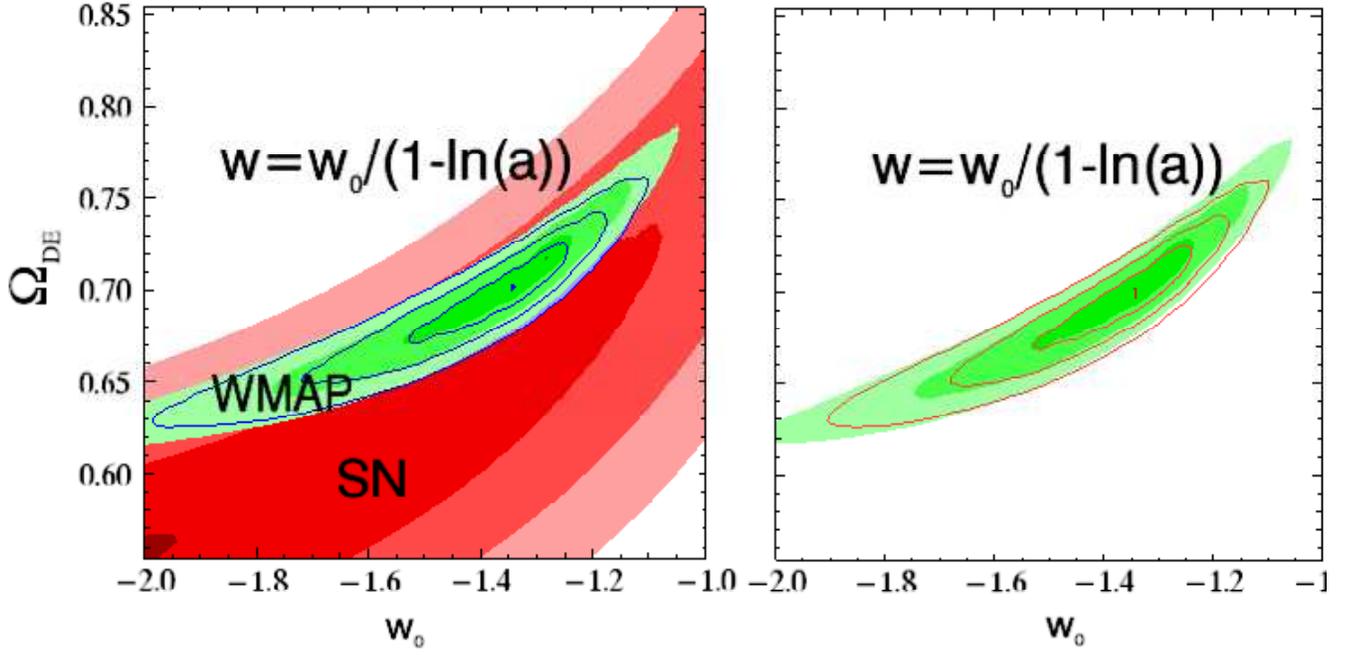}
\vspace{-15cm} \caption{ 3$\sigma$ WMAP and SN constraints on the
parametrized quintom model $w= w_0/(1-\ln a)$ shown together with
the best fit values. Left panel: Separate WMAP and SN constraints.
The green(shaded) area is WMAP constraints on models where dark
energy perturbations have been included and the blue area( contour
with solid lines) is without dark energy perturbations. Right panel:
Combined WMAP and SN constraints on the parametrized quintom model
with perturbations(green/shaded region) and without
perturbations(red region/contour with solid lines). }\label{Fig7}
\end{center}
\end{figure}

\end{document}